# HOTLink[1] Rack Monitor

A.R. Franck, R.W. Goodwin, P.A. Kasley, M.F. Shea
FNAL[2], Batavia, IL 60510

**Abstract**
A remote data acquisition chassis, called a HOTLink Rack Monitor, HRM, has been developed for use in the Fermilab control system. This chassis provides for 64 analog input channels, 8 analog output channels, and 8 bytes of digital I/O. The interface to the host VMEbus crate is by way of a 320 MHz HOTLink serial connection to a PMC mezzanine module. With no processor intervention, all data sources in the remote chassis are read at 100 µsec intervals, timestamped, and stored in a 2 MB circular buffer on the PMC module. In operation, the memory always contains the most recent 16k samples of 10 kHz data from all 64 analog input channels. An expansion module that resides in the HRM chassis records snapshot data for 8 analog channels, each channel consisting of up to 16k readings, digitized at rates up to 10 MHz. Snapshot data is also returned and stored in buffers on the PMC module. Because the HRM presents a memory-mapped interface to the host, it is independent of the operating system and may be used in any system that supports PMC mezzanine modules.

## 1 Introduction

A new data acquisition chassis has been developed for use in the Fermilab control system. The goal of this device is to collect both slow and fast analog data, along with providing analog outputs, digital input and digital output data, and to allow for future expansion. This device can be located several tens of meters from the control computer.

## 2 Design of the HRM

Although many of today's accelerator control systems use VMEbus crates and components, it is not clear that standard will continue to be used in the future. Other more cost effective platforms, such as PC hardware, are being used in some installations. To eliminate the dependence on the platform used to drive it, the HRM is structured as a remote chassis connected to a PCI mezzanine card (PMC). The PMC module can then be mounted on a PMC carrier card mounted in a VMEbus, PC chassis, CPCI crate, or other system.

The connection between the HRM chassis and the PMC module is a commercially available high-speed serial protocol called HOTLink™, a product of Cypress Semiconductor Corp. This serial link operates at either 160 or 320 Mbits/sec. The normal speed will be 320 Mbits/sec for installations where the HRM is close to the control computer; 160 Mbits/sec may be useful if the HRM chassis is placed farther away.

In control applications at Fermilab, analog signals typically fall into one of two distinct categories: slow signals and fast signals. Slow signals include dc levels, sample and hold amplifier outputs, and slowly varying signal sources. These signals are slow enough that high input impedance digitizers may be used without worrying about cable reflections. The HRM accepts slow signals into multiplexed digitizers that digitize all inputs each time they are triggered. The HRM samples all slow data channels continuously at a rate of 10 kHz– fast enough to provide information about power supply ripple, rf amplitude envelopes, etc.

Fast signals typically come from sources that can drive 50-ohm cable and need termination to avoid distortion from termination mismatch. These signals are treated as snapshot data. That is, signals are only digitized when a trigger occurs, usually to collect specific waveforms at a preselected time. The HRM has an optional eight channel, 14-bit digitizer that can record data at selectable sample rates up to 10 Msamples/sec over an input range of ± 5 V. Each of the eight channels has its own 14-bit digitizer followed by a 16K sample FIFO. Signals faster than a few MHz must be looked at using digital oscilloscopes or special data acquisition boards.

Some applications may need higher resolution snapshot data. A 16-bit 100kHz snapshot option board is being developed for these applications.

Like other Fermilab Rack Monitors, the HRM provides for 64 analog inputs, 8 analog outputs and 8 bytes of digital I/O signals. The 64 analog inputs are multiplexed to four Analog Devices AD-976, 16-bit, 200 kHz digitizers. At a repetition rate of 10 kHz, all 64 analog channels are digitized and the data is placed in

---

dual port RAM before being transmitted via HOTLink to the PMC interface module. At the same 10 kHz rate, all 8 bytes of digital I/O, the setting values of the eight analog output D-A channels, and several other control registers are all read and sent to the PMC module. The slow digitizers have 16-bit resolution; the analog outputs are derived from 16-bit DACs

The HRM chassis is controlled by an Analog Devices ADSP-2181 Digital Signal Processor. The chassis includes an analog board, a digital I/O board and a central control board that contains the HOTLink transmitter and receiver, and the memory-bus connection to the analog and digital boards. The 8-channel quick digitizer and additional future boards connect to this same memory bus.

### 3 PMC Interface Module

The control computer's access to the HRM is through the PMC module. This module includes the PCI interface circuitry, four MBytes of static RAM, Altera programmable logic chips, and the HOTLink transmit and receive circuitry. From the computer the HRM appears as a block of memory. Data is sent autonomously from the HRM remote chassis to be stored in the onboard memory. Slow A-D data is sent to this memory at 100 µs intervals and is stored in two MBytes of circular buffer. Digital I/O data, D-A settings and remote register data are also sent to the PMC module at 100µs intervals. When the computer writes data to the register memory space, the same data is transmitted to the remote HRM chassis. Similarly, data from optional boards placed in the HRM chassis will be sent to the PMC module and stored in its memory as needed with no involvement from the control computer.

The design of the HRM and its associated PMC module results in a data acquisition system that requires very little involvement from the control computer. After the HRM chassis is enabled by setting a single bit in a control register, analog and digital data stream into the PMC RAM at 100 µs intervals, with no further involvement of the control processor. D-A and digital output settings are made by storing a word or byte of data in the register area of PMC memory. These digital settings are made immediately, but there is a latency of up to 100 µS before confirmation of the success of the setting will be stored in the PMC RAM. Although slow analog data may be easily read from PMC memory, the computer does need to know how to access the most recent or the particular range of data stored in the circular buffer.

Each 16 bit word of data sent between the PMC module and the HRM chassis is sent as a 5-byte transmission; one byte of packet header (K28.0 Special Character) followed by 4 information bytes.. The first two bytes comprise the 16 bit data field. The third and fourth bytes carry 12 bits of channel number or other address and a 4-bit Type Code that indicates to the receiver the type of data contained in the packet. Two contiguous packets are used to send 32-bit data, and one bit of the addressing is used to indicate whether the data is the MS or LS word. Each packet stands on its own – no other context is needed to determine how to handle it at the PMC.

These packets are strung together to form a return frame that has recurring and non-recurring parts. Digital I/O and register updates are placed at the front of the frame, followed by slow-data. These elements are sent in every 100us frame. The remainder of the frame is filled out with snapshot data.

### 4 Timestamping the data

At Fermilab, timing information is sent to accelerator devices using the Tevatron Clock, (TCLK), a 10 MHz modified Manchester clock encoded with 8-bit timing events. Wherever timing is needed, TCLK is received, events are decoded and delayed triggers are generated for use by accelerator devices. To plot analog data from various sources on the same graph, all data must be timestamped to allow it to be referenced to TCLK. Front end control computers receive TCLK and record the occurrence of encoded events by storing the event number along with the contents of a free-running 32-bit, 1 MHz counter. The HRM chassis has its own TCLK receiver and programmable event decoder.

The HRM chassis captures two types of timestamps. A slow-data timestamp is captured from a free-running, 32-bit, 1 Mhz counter at the start of each 100us interval. This timestamp precedes the slow data portion of the return frame. A 32-bit snapshot timestamp is captured from the same counter when a preselected clock event or an external signal triggers a snapshot acquisition. This snapshot timestamp is returned to the PMC at the very beginning of the snapshot data

The timestamps returned to the PMC module will have an offset from similar timestamp values measured by the front end control computer, but this offset can be determined at each occurrence of the selected TCLK event. With this offset, the control computer can relate timestamp readings to TCLK. Because the timestamps are collected by the hardware and are not dependent on software response times, timing of data from a given HRM is known to within ±1µs.

### 5 Data Storage

Table 1 shows the memory map for the PMC module. D-A setting readbacks, digital I/O data and other register data from the HRM are all stored in fixed memory blocks on the PMC module. Snapshot data from the quick digitizers are also stored in known locations. Slow analog data storage is more involved because it is stored as 64-channel data blocks in a 2 MByte circular buffer. At any given time, the circular buffer contains 16,000 data sets representing the values of all channels for the past 1.6 seconds. Old data is overwritten by new data as it is received.

Each 100µs interval, the HRM transmits a 32-bit timestamp followed by the 64-channel block of analog data values. The arrival of the timestamp increments a block counter on the PMC module. The value of this block counter is then used to form the address of a location in a circular buffer where the timestamps are stored. Slow analog data following the timestamp are stored in a second circular buffer whose address is also formed using the block counter value.

In normal operation, the control computer accesses HRM register data as specific locations in PCI memory space. Front end computers keep a data pool of analog values refreshed at regular intervals, usually 15 Hz. To access the most recent analog values, the control computer reads the block counter and forms the address of the block of data that is currently or was most recently filled with new data. These values, or the values of the preceding block of data, are copied into the data pool for use by other programs.

Snapshot data from the fast digitizer board are stored in a fixed 256KB area of memory with data from each channel stored in a contiguous 32KB block. The fast data timestamp is stored in a fixed location. A counter on the PMC module increments as fast data is received, and its value is used to form the address where the data is to be stored. It is also used by the front end computer to determine how much data has been received.

The fast digitizer operates in one of two modes, single or repetitive. In the single mode, the front end computer must arm the digitizers to enable them to respond to the next trigger. The repetitive mode allows the fast digitizers to respond to each trigger. In this mode, the FIFO is reset and new data is stored whenever a trigger is received, even if the digitizers were still storing data from the previous trigger.

Table 1: Memory map for the PMC module

| Memory Space | Starting Address | Range | Type Code |
| --- | --- | --- | --- |
| **Slow Data** | $000000 | 2 MB | 0 |
| **Option Board Data** | $200000 | 2 MB | 1 |
| **Snapshot Data** | $400000 | 256 KB | 4 |
| **Snapshot Timestamp** | $45FFFC | 4 Bytes | 5 |
| **Slow Data Timestamp** | $460000 | 64 KB | 6 |
| **Option Bd Timestamp** | $470000 | 64 KB | 7 |
| **Remote Registers** | $480000 | 8 KB | C…F |
| **Error and Status Reg** | $500000 | 1 MB | na |

## 6 Packaging

The HRM is a self-powered, 2U, 19 inch rack mounted chassis. It connects to its associated PMC module by two twisted pair cables used for the transmit/receive data path. The PMC module can be mounted on a PMC carrier board or on a PMC socket of the front end processor board when available. Within the HRM chassis, there is a central board which holds the DSP, the HOTLink transmitter and receiver, and interfaces to the various I/O boards (digital, slow analog, 8-channel 10 MHz quick digitizer board) HOTLink speed is cable selectable. With the HOTLink operating at 320 Mbits/sec, transmission of the slow analog data uses 10µS of the 100 µs frame time, which is only 10% of the capacity of the HOTLink. For cases where the link is operating at 160Mbits/sec, the link utilization rises to 20%.

## 7 Status of the HRM project

At the present time, prototypes of the HRM chassis and the PMC interface module have been tested and are operational. A final version of the circuit boards needs to be developed to incorporate the latest version of the DSP onto the central board, to implement a fly-by DMA bus to better accommodate possible future increases in serial link speed, and to simplify I/O card mounting and chassis assembly. After that is finished, the HRM will be used for a number of data acquisition tasks within the Fermilab controls department. It is planned that this design will be offered to commercial vendors to fabricate and market as a product.

## 8 Conclusions

The HRM provides a cost effective way to interface accelerator equipment to the control system. From the software standpoint, the HRM requires very little support from the control computer. Once enabled, data independently returns to the PMC module where it becomes available as values stored in normal PCI memory space. There is no dependence on the particular processor or operating system used in the front end. The DSP is self-contained and its software does not require downloading by the control computer. These features make it easy to integrate the HRM into any control system. Multiple HRMs can be added to a single front end computer simply by adding PMC interface modules. Future enhancements of the HRM are possible by designing additional data acquisition boards and by using the DSP to preprocess raw data before it is shipped to the PMC.